\documentclass[aps,prl,showpacs,superscriptaddress,floatfix,onecolumn,preprint]{revtex4-1}
\usepackage{graphicx}
\usepackage{dcolumn}
\usepackage{bm}
\usepackage{setspace}
\usepackage{psfrag}
\usepackage{subfigure}
\usepackage{amsfonts}
\usepackage{braket}
\usepackage{color}
\usepackage{pdfpages}

\begin{document}

\title{Spin polarized electrons produced  by strong field ionization}
\author{Ingo Barth}
\author{Olga Smirnova}
\affiliation{Max Born Institute, Max-Born-Str. 2A, 12489, Berlin, Germany}
\begin{abstract}
We show that ionization of noble gas atoms by a
strong infrared circularly polarized laser field under standard experimental conditions
can yield electrons with up to 100\% spin polarization in energy resolved measurements. Spin polarization arises
due to the interplay of the electron-core entanglement and  the
sensitivity of ionization in circularly polarized fields
to the sense of electron rotation in the initial state.
\end{abstract}
\maketitle

Coherent ultrashort light \cite{asec_pulses} and electron beams \cite{asec_pulses,hommelhof} produced during the interaction of atoms,
 molecules and solids with strong infrared laser fields are promising new tools for ultrafast spectroscopy.
Photoelectrons extracted by the strong laser field from the metal nano-tip  can form intense, few tens of femtosecond long coherent
electron pulses \cite{hommelhof}, opening new opportunities for ultrafast electron diffraction within a table top setup.
Photoelectrons produced via strong field ionization of atoms and molecules can serve as an attosecond probe of optical tunneling
  \cite{ursi1,ursi2,corkum1,ursi3}, molecular structure
 \cite{science_paul, jphysb_olga} and dynamics \cite{CDlin}; their coherence can be  used to record holographic
 images of atomic core \cite{sciece_huismans,jphysb_olga}. We show that, when produced by ionization in
a strong infrared circularly polarized field under standard experimental conditions \cite{ursi1,ursi2,corkum1,ursi3},
coherent ultrashort  photoelectron pulses can
have high and controllable degree of spin polarization, opening
new opportunities for attosecond spectroscopy.

Analysing one-photon ionization, U. Fano \cite{Fano} has shown that usually
weak effects of the spin-orbit interaction  are strongly enhanced in the vicinity of
the Cooper minima in the photoionization continua,
leading to 100\% spin polarization within a certain energy window.
In the one photon ionization, spin polarization can also be achieved via
ionization from a particular fine structure level of an atom or a molecule \cite{cherepkov}.
Elegant extension to resonant multi-photon ionization in the weak-field (perturbative)
limit has been proposed by P. Lambropoulos \cite{Lambropoulos1,Lambropoulos2,Lambropoulos3}. Importantly, it has been
demonstrated that high degree of spin-polarization is not always associated with minima in cross sections \cite{Lambropoulos4,Lambropoulos5}. 
For example, 100\%  spin polarization is achieved away from the minimum in the three-photon ionization cross section of alkali atoms \cite{Lambropoulos4}, 
and at the maximum of the one-photon cross section for Xe \cite{Lambropoulos5}.

All of these mechanisms rely on fine tuning the light frequency and require
long, low-intensity pulses.
In contrast, our mechanism does not
rely on frequency tuning or intermediate resonances. It operates in the strong-field
regime, for broad range of frequencies and for short pulses.
Spin polarization is achieved via spin-orbit interaction in the ionic core and is
due to the interplay of (i) the electron-core entanglement and (ii) the
sensitivity of ionization in circularly polarized fields
to the sense of electron rotation in the initial state.

Consider strong field ionization of noble gas atoms by right circularly polarized field
propagating in the positive direction of the $z$-axis.
For all noble gas atoms except Helium, the outer shell is filled by six $p$ electrons.
Thus,  there is no spin-orbit interaction in the ground state and there is equal amount of $p_+$ and $p_-$ electrons, 'counter-rotating'  and 'co-rotating'   with the field.
We have recently shown  \cite{PRA1,PRA2} that non-adiabatic effects in strong-field ionization
 result in its high sensitivity  to the sense of
electron rotation in the initial state: circularly polarized infrared laser field preferentially removes counter-rotating electrons.
Our theoretical prediction has now been confirmed by the experiment \cite{herath}.

Electron removal leaves the $p$-shell open. Spin-orbit interaction splits
the states of the ion  with respect to the total angular momentum of the core  $J=1/2$ and $J=3/2$,
providing two ionization channels with slightly different ionization potentials.
Just like in the EPR experiment, once we divided the system with total angular momentum $J=0$ , $L=0$ and $S=0$ into two parts electron and the core,
 we know that $m_j=-M_J$, $m_s=-M_s$, where capital letters and small letters are for  the initial values of the quantum numbers for the core and the electron correspondingly.

Ionization of $m_l=0$ is strongly suppressed.  Right circular field preferentially removes $p^-$ electron  \cite{PRA1,PRA2}, thus, $m_l =-1$. 
Suppose we have created the ion in $^2P_{1/2}$ state: for the core $J=1/2$, $|M_J|=1/2$, $L=1$ and therefore $M_L$ and $M_S$ have the opposite sign.
 Thus, the $p^-$ electron correlated to the state $^2P_{1/2}$ must have had $|m_j|=1/2$ and $m_l$ opposite to $m_s$, yielding $m_s=1/2$ for the initial value $m_l =-1$.
 Thus, the interplay of electron-ion entanglement and sensitivity of ionization to initial $m_l$ lead to spin-polarization.
100\% selectivity of ionization to the sense of rotation  of the electron in the ground state would lead
to 100\% spin polarization in the channel $^{2}P_{1/2}$. Spin polarization in the channel $^{2}P_{3/2}$ is less than 100 \%, since the total momentum of the core $J=3/2$ admits
both $|m_j|=3/2$ and $|m_j|=1/2$ of the correlated photoelectron.
The ability to separate  photoelectron spectra corresponding to $^{2}P_{3/2}$  and  $^{2}P_{1/2}$ ionization channels experimentally \cite{rottke}
offers opportunities for achieving high degree of spin polarization of coherent electron beams produced by strong field ionization.
Note that similar separation of strong field photoelectron spectra correlated to different core states of a polyatomic molecule has recently been demonstrated in \cite{albert} and used to identify different channels in strong field ionization.

To provide quantitative picture of the effect, we extend our method \cite{PRA1,PRA2} to include spin-orbit interaction.
The extension is based on angular momentum algebra and does not contain any further approximations. Pertinent theoretical work in case of linearly polarized fields includes \cite{santra,Kornev}.

Nonadiabatic ionization rates for atomic $p_m$ orbitals ($m=0,\pm 1$) in left ($c=-1$)
or right ($c=+1$) circularly polarized laser fields
can be written as a sum over multiphoton channels \cite{PRA1,PRA2}
\begin{eqnarray}
\label{multiphoton}
w_c^{p_m}(\mathcal{E},\omega,I_p)&=&\sum_{n\geq n_0}^\infty w_{nc}^{p_m}(\mathcal{E},\omega,I_p),
\end{eqnarray}
where \ $n_0=(2U_p+I_p)/\omega$.
Summation leads to the following simple expression \cite{PRA1,PRA2}:
\begin{eqnarray}
\label{ionpm}
w_c^{p_m}(\mathcal{E},\omega,I_p)&=&|C_{\kappa l=1}|^2I_p\,\frac{\mathcal{E}}{2\mathcal{E}_0}\,h_c^{p_m}(\gamma)e^{-\frac{2\mathcal{E}_0}{3\mathcal{E}}\,g(\gamma)}.
\end{eqnarray}
In Eqs.\,(\ref{multiphoton}) and (\ref{ionpm}), $\mathcal{E}$ is the electric field amplitude, $\omega$ is the laser frequency, $I_p$ is the ionization potential,
$U_p=\mathcal{E}^2/(4\omega^2)$ is the pondermotive potential,
$\mathcal{E}_0=(2I_p)^{3/2}$,
$\gamma=\sqrt{2I_p}\,\omega/\mathcal{E}$ is the Keldysh parameter \cite{keldysh}. The coefficient
$C_{\kappa l=1}$ characterizes the asymptotic behavior of the radial wave function, depending on $\kappa=\sqrt{2I_p}$ and the
orbital quantum number $l$, with $l=1$
for $p_m$ orbitals. The  exponential factor  $g(\gamma)$ \cite{PRA1,PRA2}
does not depend on the sense of circular polarization $c=\pm1$ and on the parameters of atomic orbital.
The orbital dependence comes from the prefactors $h_c^{p_0}(\gamma)$ and $h_c^{p_\pm}(\gamma)$ for $p_0$ and $p_\pm$ orbitals as
shown in Refs.\,\cite{PRA1,PRA2} and results in
higher ionization rates for $p_-$ orbitals than  for $p_+$ orbitals  in right circularly polarized laser fields ($c=+1$).
\begin{widetext}
\begin{figure}
\begin{center}
 \includegraphics[width=0.5\textwidth]{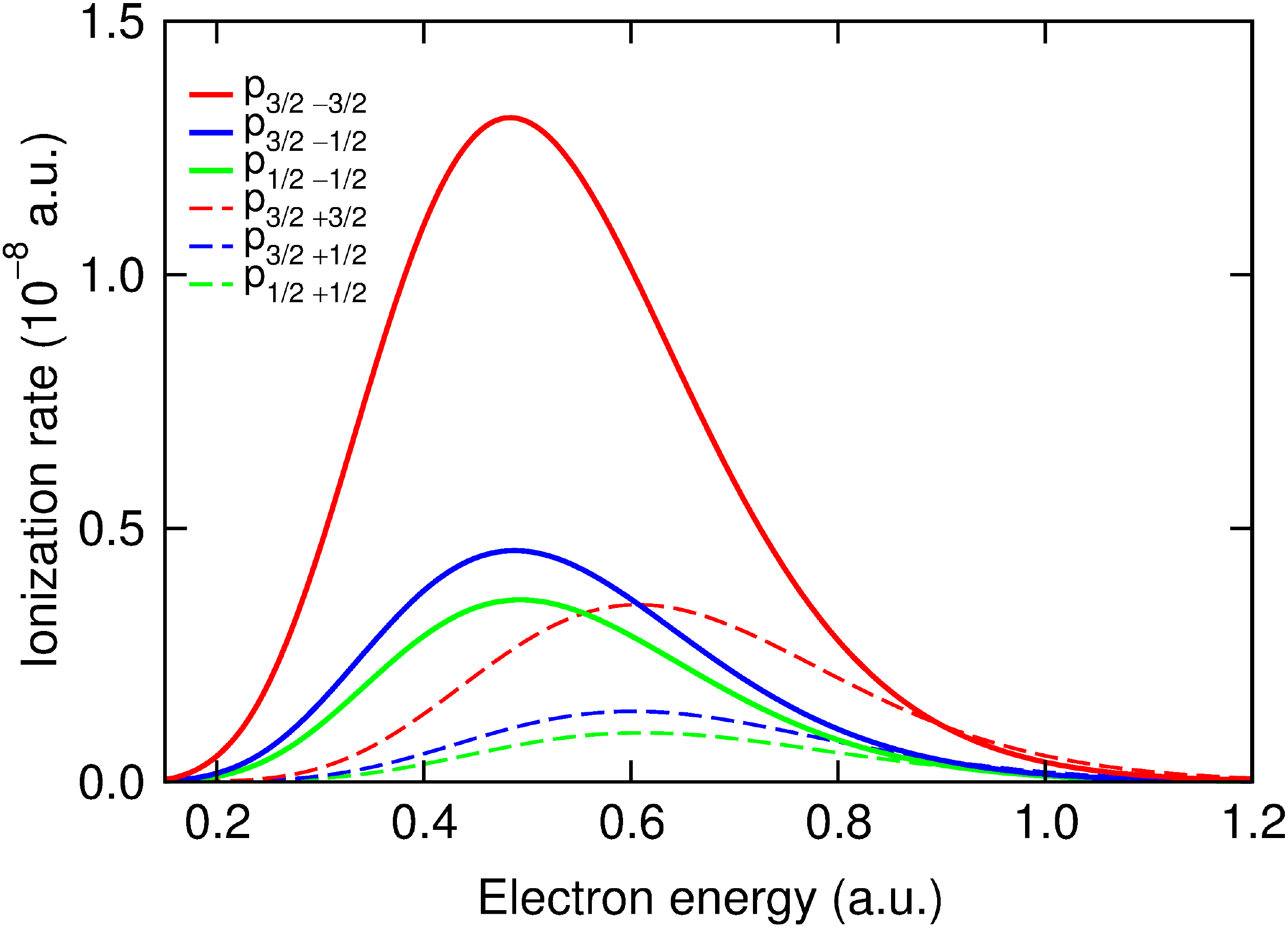}\includegraphics[width=0.5\textwidth]{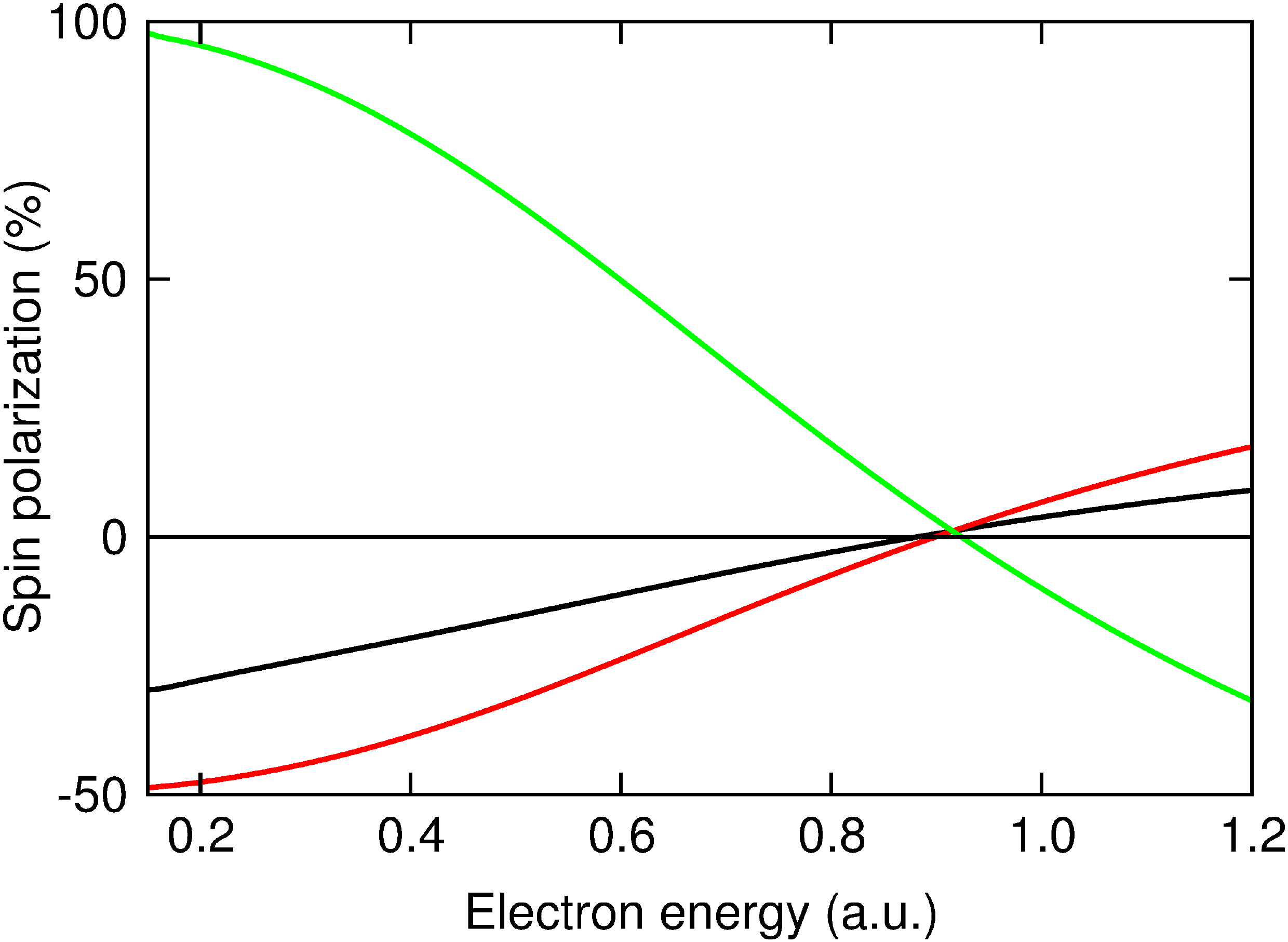}
\caption{Left panel shows photoelectron energy distribution  (Eq.\,(\ref{ionpjmn})) for $p_{jm_j}$ spin-orbitals.
Right panel shows spin polarization of  photoelectrons  (Eq.\,(\ref{spinpoln})) resolved on $^2P_{1/2}$ state of the core (green curve),
 $^2P_{3/2}$ state of the core (red curve) and integrated over core states (black curve),for krypton atom, ionization potentials $I_p^{P_{3/2}}=0.5145\,$a.u.\ and $I_p^{P_{1/2}}=0.5389\,$a.u.
and right circularly polarized field with frequency $\omega=0.057$\,a.u.\ (800\,nm) and field strength $\mathcal{E}=0.05$\,a.u. ($1.8\cdot10^{14}$ W/cm$^2$)}
\end{center}
\end{figure}
\end{widetext}

The relationship between the ionization rates from the non-relativistic orbitals considered above and the relativistic spin-orbitals
 $p_{jm_j}$ with total (orbital and spin) angular quantum number $j$ and a corresponding magnetic quantum number $m_j$ obtains using angular momentum algebra.
 The orbitals $p_{jm_j}$
can be expanded in the basis of the products of orbitals $p_m$ and spin functions $\chi_{sm_s}$ as
\begin{eqnarray}
\label{expansion}
p_{jm_j}&=&\sum_{m,m_s}C^{jm_j}_{1m,\frac12m_s}p_m\chi_{\frac12 m_s},
\end{eqnarray}
where the expansion coefficients $C^{jm_j}_{lm,sm_s}$ are the Clebsch-Gordan coefficients with
orbital and spin quantum numbers $l=1$ and $s=1/2$, respectively;
the corresponding magnetic quantum numbers $m$ and $m_s$ are restricted by $m+m_s=m_j$.
Integrating the corresponding density over the spin variable $\sigma$ yields the orbital density of spin-orbitals $p_{jm_j}$
\begin{eqnarray}
\label{density}
\int |p_{jm_j}|^2d\sigma&=&\sum_{m,m_s}\left|C^{jm_j}_{1m,\frac12m_s}\right|^2|p_m|^2.
\end{eqnarray}
The same relations hold for the momentum representation of spin-orbitals $\tilde p_{jm_j}$, i.e.
\begin{eqnarray}
\label{densitymomentum}
\int |\tilde p_{jm_j}|^2d\sigma&=&\sum_{m,m_s}\left|C^{jm_j}_{1m,\frac12m_s}\right|^2|\tilde p_m|^2,
\end{eqnarray}
where $\tilde p_m$ is the momentum representation of orbitals $p_m$.
Since the strong-field ionization rates (Eqs.\,(\ref{multiphoton}) and (\ref{ionpm})) depend linearly on $|\tilde p_m|^2$ (see Refs.\,\cite{PRA1,PRA2}),
we can express the ionization rate for the spin-orbitals $ p_{jm_j}$ (Eq.\,(\ref{densitymomentum})) via ionizaton rates  for $ p_m$ orbitals:
It yields the general formula for the ionization rates for $p_{jm_j}$ spin-orbitals
\begin{eqnarray}
\label{ionpjm}
w_c^{p_{jm_j}}(\mathcal{E},\omega,I_p^{P_j})&=&\sum_{m,m_s}\left|C^{jm_j}_{1m,\frac12m_s}\right|^2 w_c^{p_m}(\mathcal{E},\omega,I_p^{P_j}),
\end{eqnarray}
in particular
\begin{eqnarray}
\label{p1212}
w_c^{p_{\frac12\pm\frac12}}(\mathcal{E},\omega,I_p^{P_\frac12})&=&\frac23\,w_c^{p_\pm}(\mathcal{E},\omega,I_p^{P_\frac12})+\frac13\,w_c^{p_0}(\mathcal{E},\omega,I_p^{P_\frac12}),\\
\label{p3212}
w_c^{p_{\frac32\pm\frac12}}(\mathcal{E},\omega,I_p^{P_\frac32})&=&\frac13\,w_c^{p_\pm}(\mathcal{E},\omega,I_p^{P_\frac32})+\frac23\,w_c^{p_0}(\mathcal{E},\omega,I_p^{P_\frac32}),\\
\label{p3232}
w_c^{p_{\frac32\pm\frac32}}(\mathcal{E},\omega,I_p^{P_\frac32})&=&w_c^{p_\pm}(\mathcal{E},\omega,I_p^{P_\frac32}),
\end{eqnarray}
or as a sum over multiphoton channels
\begin{eqnarray}
w_c^{p_{jm_j}}(\mathcal{E},\omega,I_p^{P_j})&=&\sum_{n\geq n_0}^\infty w_{nc}^{p_{jm_j}}(\mathcal{E},\omega,I_p^{P_j}),
\end{eqnarray}
where
\begin{eqnarray}
\label{ionpjmn}
w_{nc}^{p_{jm_j}}(\mathcal{E},\omega,I_p^{P_j})&=&\sum_{m,m_s}\left|C^{jm_j}_{1m,\frac12m_s}\right|^2 w_{nc}^{p_m}(\mathcal{E},\omega,I_p^{P_j}).
\end{eqnarray}
Here, $I_p^{P_J}$ is the $J$-dependent ionization potential due to spin-orbit splitting between $^2P_{1/2}$ and $^2P_{3/2}$ states of the ion.
The quantum number $m_s=\pm1/2$ in equations (\ref{ionpjm}, \ref{ionpjmn}) indicate projection of the electron spin on the laser propagation direction.
Thus, equations (\ref{ionpjm}, \ref{ionpjmn})  provide information on spin-resolved ionization rates.
Total spin polarization is proportional to the difference in  the
total ionization rates for the photoelectrons with spin up $w_{c\uparrow}(\mathcal{E},\omega)$ and down $w_{c\downarrow}(\mathcal{E},\omega)$ :
\begin{eqnarray}
\label{spinpol}
P_c(\mathcal{E},\omega)=\frac{w_{c\uparrow}(\mathcal{E},\omega)-w_{c\downarrow}(\mathcal{E},\omega)}{w_{c\uparrow}(\mathcal{E},\omega)+w_{c\downarrow}(\mathcal{E},\omega)}.
\end{eqnarray}
Using Eq.\,(\ref{ionpjm}), total spin-resolved ionization rates can be expressed via $m$-resolved ionization rates $w_c^{p_m}$:
\begin{eqnarray}
\label{wcu}
w_{c{\uparrow},{\downarrow}}(\mathcal{E},\omega)&=&\frac13\,w_c^{p_0}(\mathcal{E},\omega,I_p^{P_\frac12})+\frac23\,w_c^{p_0}(\mathcal{E},\omega,I_p^{P_\frac32})+w_c^{p_\pm}(\mathcal{E},\omega,I_p^{P_\frac32})\\\nonumber
&&+\frac23\,w_c^{p_\mp}(\mathcal{E},\omega,I_p^{P_\frac12})+\frac13\,w_c^{p_\mp}(\mathcal{E},\omega,I_p^{P_\frac32}),
\end{eqnarray}
where the upper superscript in $w_c^{p_\pm}$  should be used for spin-up $(\uparrow)$ and the lower superscript should be used for spin-down ($\downarrow$) rates correspondingly.

In particular, neglecting  small contribution of $w_c^{p_0}$ \cite{PRA2} in Eq.(\ref{wcu}), yields simple and accurate expressions for spin-polarization of electron correlated to states $^2P_{1/2}$:
\begin{eqnarray}
\label{spinpol1}
P_c(\mathcal{E},\omega,I_p^{P_\frac12})
&\simeq&2\,\mathrm{sgn}(c)\,\frac{A(\gamma)}{1+A(\gamma)^2},
\end{eqnarray}
and $^2P_{3/2}$:
\begin{eqnarray}
\label{spinpol2}
P_c(\mathcal{E},\omega,I_p^{P_\frac32})
&\simeq&-\mathrm{sgn}(c)\,\frac{A(\gamma)}{1+A(\gamma)^2},
\end{eqnarray} where
\begin{eqnarray}
A(\gamma)&=&\frac{\zeta_0}{\gamma}\sqrt{\frac{1+\gamma^2}{\zeta_0^2/\gamma^2+1}}.
\end{eqnarray}
Here $\gamma $ is the Keldysh parameter and the parameter $0\leq\zeta_0\leq1$ satisfies the equation $\sqrt{\frac{\zeta_0^2+\gamma^2}{1+\gamma^2}}=\tanh\frac{1}{1-\zeta_0}\sqrt{\frac{\zeta_0^2+\gamma^2}{1+\gamma^2}}$. Note that  $\zeta_0\simeq\gamma^{2}/3$ for $\gamma\ll1$, and $\zeta_0\simeq1-1/\ln\gamma$ for $\gamma\gg1$ \cite{PPT1}.
Effects of long-range potential  equally affect the ionization rates $w_c^{p_\pm}$, $w_c^{p_0}$ \cite{PRA1,PRA2} and thus they do
not affect results for spin-polarization given by Eqs.(\ref{spinpol1},\ref{spinpol2}) .

Prior to the analysis of total spin polarization,
it is essential to consider spin polarization resolved on the
final electron energy and the final state of the core. Note, that the latter is easily accomplished
 by energy discrimination of the photoelectron spectra correlated to different core states as in \cite{rottke,albert}.
It is obtained using Eq.\,(\ref{ionpjmn}):
\begin{eqnarray}
\label{spinpoln}
P_{nc}(\mathcal{E},\omega,I_p^{P_J})=\frac{w_{nc\uparrow}(\mathcal{E},\omega,I_p^{P_J})-w_{nc\downarrow}(\mathcal{E},\omega,I_p^{P_J})}
{w_{nc\uparrow}(\mathcal{E},\omega,I_p^{P_J})+w_{nc\downarrow}(\mathcal{E},\omega,I_p^{P_J})}.
\end{eqnarray}
where the corresponding rates $w_{nc\uparrow}(\mathcal{E},\omega,I_p^{P_J})$ and $w_{nc\downarrow}(\mathcal{E},\omega,I_p^{P_J})$ are resolved on
the number of absorbed photons, i.e. on the final electron energy: $E_{kin} = (n-n_0)\omega$ \cite{PRA1,PRA2}.
Energy and spin-resolved ionization rates $w_{nc\uparrow,\downarrow}(\mathcal{E},\omega,I_p^{P_J})$
 are expressed via energy and $m$-resolved ionization rates $w_{nc}^{p_m}$  given by Eq. (\ref{multiphoton})
in the same way as in Eq.\,(\ref{wcu}).
\begin{figure}[t]
\includegraphics[width=0.5\textwidth]{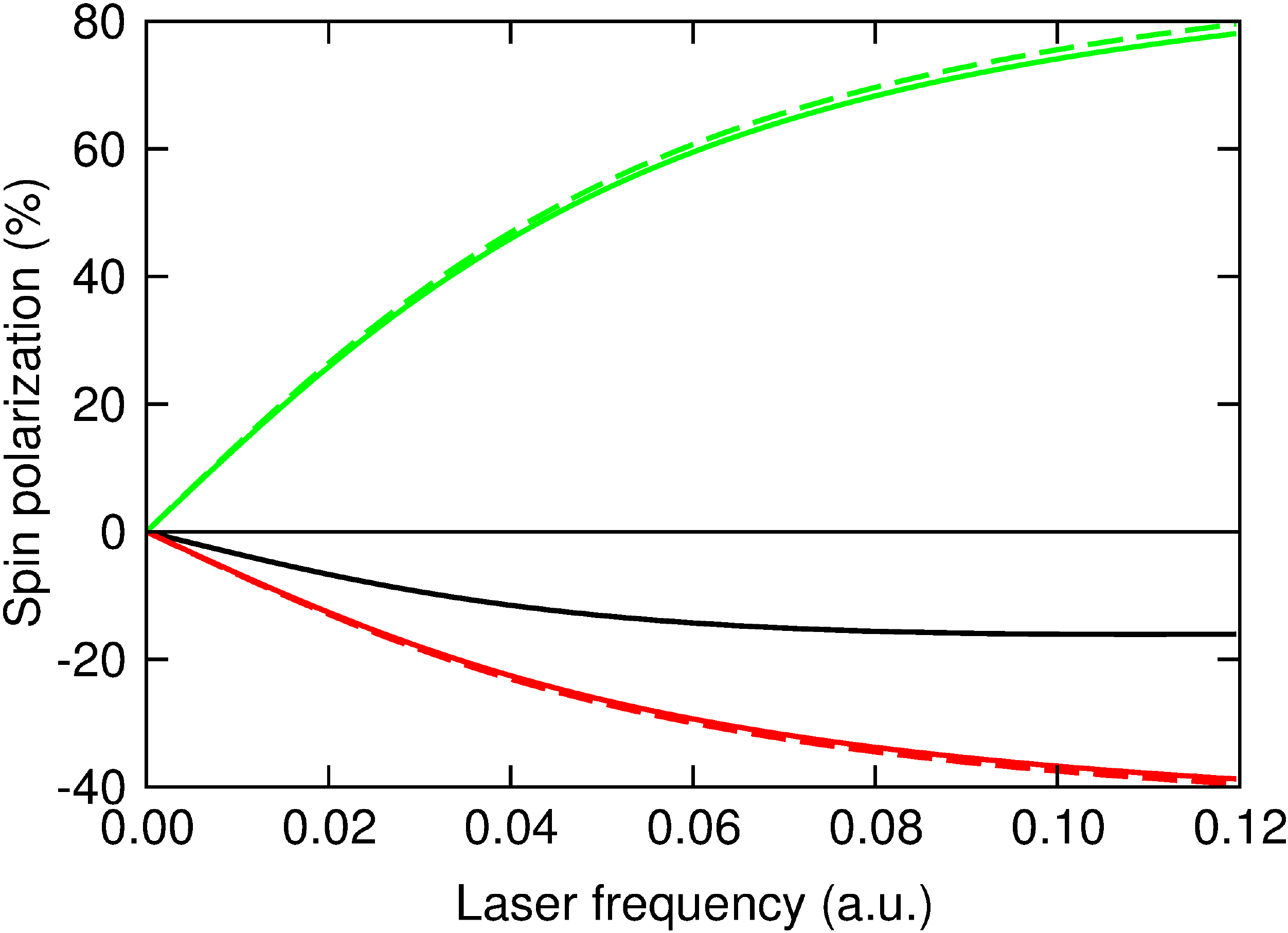}\includegraphics[width=0.5\textwidth]{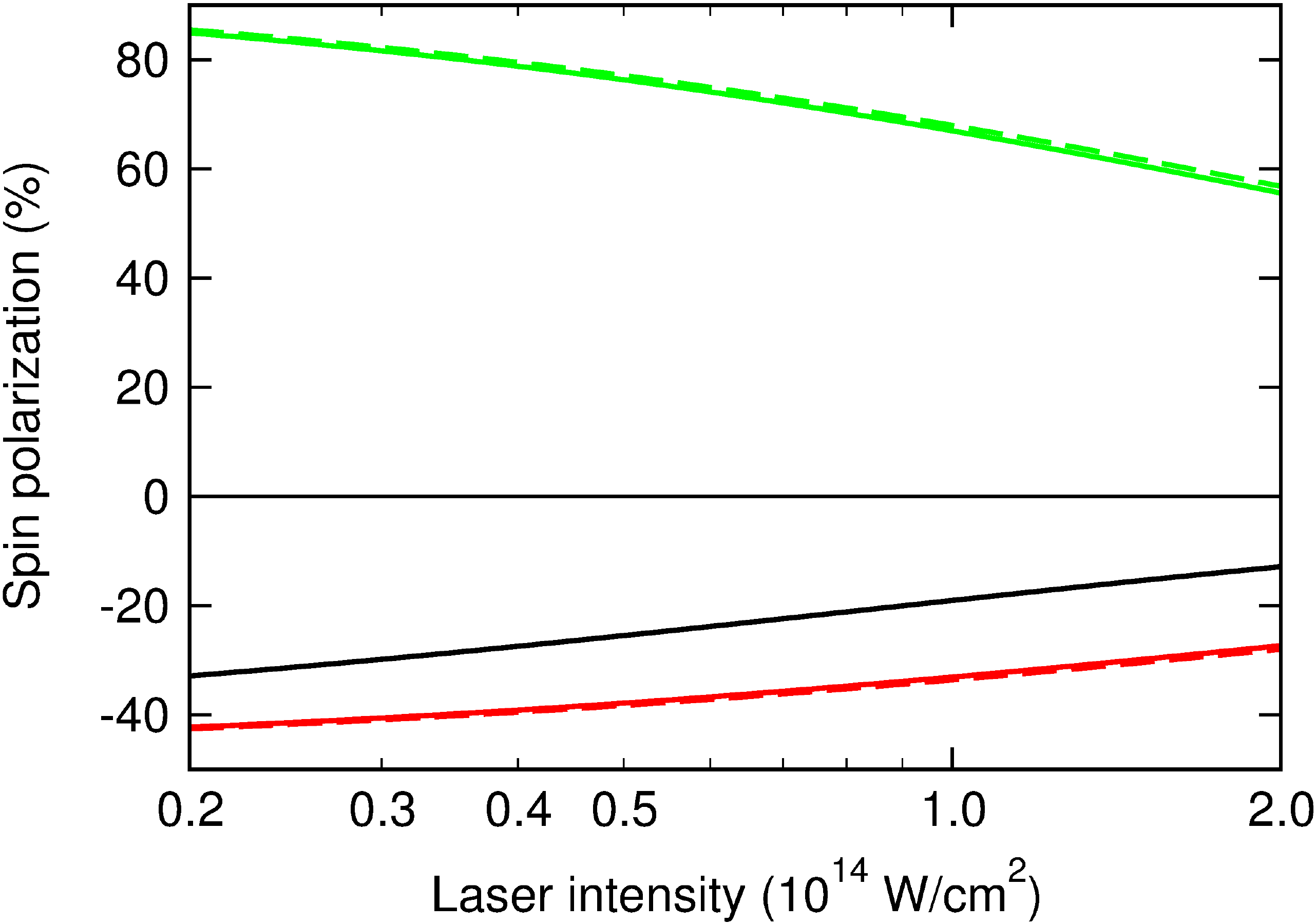}
\caption{Control of spin polarization in strong field ionization of Kr (ionization potentials $I_p^{P_{3/2}}=0.5145\,$a.u.\ and $I_p^{P_{1/2}}=0.5389\,$a.u.) by right circularly polarized laser field.
 Energy-integrated spin polarization resolved on $^2P_{1/2}$ state of the core (green curve- accurate, green dashed curve- approximate using Eq.\ref{spinpol1}),  $^2P_{3/2}$ state of the core (red curve-accurate, red dashed curve- approximate using Eq.\ref{spinpol2}) and integrated over core states (black curve). Left panel shows dependence on laser frequency, for the field strength $\mathcal{E}=0.05\,$a.u.
Right panel shows dependence on the laser intensity, for the laser frequency $\omega=0.057\,$a.u. (800\,nm).}
\end{figure}

Energy and core-state resolved photoelectron spectra for Kr atom
are shown in Fig. 1(a) for $\omega=0.057$ a.u. and $\mathcal{E}=0.05$ a.u.
corresponding to 800 nm light with intensity $1.8\cdot10^{14}$ W/cm$^2$.
Solid and dashed lines represent contributions of counter-rotating and co-rotating electrons,
resolved on the core states.
The signals coming from co-rotating and counter-rotating electrons are spectrally shifted,
reflecting non-adiabatic nature of strong-field ionization \cite{PRA1,PRA2} for these typical laser parameters.
The counter-rotating electrons dominate the low-energy part of the spectrum,
whereas the co-rotating electrons dominate the high energy part of the spectrum.

Consider first the electrons correlated  to the $^2P_{1/2}$ state of the core (green curves in Fig. 1(a)).
As discussed above, for the $^2P_{1/2}$  states, the sense of electron rotation uniquely
maps into the spin state: the solid green curve corresponds to the spin-up electrons,
while the  dashed green curve corresponds to the spin-down electrons.
The signal coming from the counter-rotaing electron (solid green curve) is much
stronger than the signal from the co-rotating electron in the low-energy part of the spectrum,
leading to high, close to 100 \% spin polarization (see Fig.1 (b), green curve) in the low energy part of the spectrum.
Since photoelectron peak correlated to  $^2P_{1/2}$ state is lower in energy than the one for the $^2P_{3/2}$ state,
separating lowest-energy electrons is efficient method of obtaining 100\% spin polarization in strong-field
ionization.

Consider now  electrons correlated to the core state $^2P_{3/2}$. Had one-to-one mapping between
the sense of electron rotation and the orientation of its spin existed for this core state,
the respective contribution of the spin-up electron would have been given solely by the red dashed curve, of the spin-down
electrons solely by the red solid curve. Thus, the spin polarization would have been given by a
curve similar to the green curve in Fig.1 (b), only with the opposite sign.

However, for the core state $^2P_{3/2}$  the picture becomes more complex, because the total momentum $J=3/2$ has more projections on the $z$-axis: $|M_J|=3/2$ and $|M_J|=1/2$.
Each sense of electron rotation in the initial state ($m$) can pair with both projections of the electron spin $m_s$.
For example, the counter-rotating electron,
 which dominates the overall signal, can have not only spin-down component (red solid curve), but also spin-up (blue solid curve) component.
Naturally, for the counter-rotating electron the spin-down component (solid red curve)  dominates the spin-up (blue solid curve) component,
 since $J=3/2$ has larger probability to have maximal possible value of the projection $|M_J|=3/2$ (i.e. larger Clebsch-Gordan coefficient).
For the same reason, the spin-up component of the co-rotating electron (red dashed curve) dominates its spin-down component (blue dashed curve).
The interplay of these four spectra  is responsible for decreased spin polarization for
the electron correlated to the $^2P_{3/2}$ core state (compare Fig. 1(b), red curve vs Fig. 1(b), green curve).

Electrons correlated to different core states have spin polarization of opposite sign and therefore
the total energy-resolved spin polarization, integrated over the two core states (Fig. 1(b), black curve), is even lower.
Such integral energy-resolved spin polarization is particularly relevant when
spectral peaks corresponding to different core states can not be resolved, e.g.
for short laser pulses.

Spin polarization in strong-field ionization is a manifestation of the  non-adiabatic nature of the process.
It vanishes in the limit of small Keldysh parameter \cite{keldysh}  $\gamma$ , when ionization rates for
co-rotating and counter-rotating electrons become equal \cite{PRA1,PRA2}.
Non-adiabaticity increases with increasing $\gamma$, offering opportunities for controlling
spin polarization of electron beams.
Fig. 2 (a,b) shows the degree of
spin polarization integrated over the final electron
energy and illustrates opportunities for its frequency and intensity control.
The degree of spin polarization can be particularly well manipulated via frequency control.

As opposed to state-resolved spin polarization, integrated spin polarization naturally depends on the strength of spin-orbit interaction.
For example, for laser frequency $\omega=0.057\,$a.u. (800\,nm),
 laser amplitude $\mathcal{E}=0.05\,$a.u. and right circular polarization, the degree of energy and core state integrated spin polarization
is $-14.0\%$ for krypton, $-20.2\%$ for xenon, and $-25.4\%$ for radon. Note that, due to the exponential sensitivity
of strong field ionization to the ionization potential, large spin-orbit splitting leads to
the suppression of ionization in channel $^2P_{1/2}$. In this case,
total spin polarization is given by Eq. \ref{spinpol2} and can never exceed 50\%.

Our work opens several new opportunities.

First, application of strong laser fields provides the opportunity to create short, dense spin-polarized electron and ion beams by using few tens of femtoseconds pulses.
Short  electron pulses could be interesting for time-resolved electron diffraction experiments. Development of femtosecond electron diffraction with coherent, ultrashort, single electron wave packets is a new direction in ultrafast spectroscopy \cite{hommelhof,Ropers,ralf}. One of the options involves an optical pump-electron probe scheme. Near IR and mid-IR laser pulse are used to generate single electron wave packets from a nano-sized particles: metallic tip, or dielectrics such as e.g. carbon or silicon nano-particles. In the latter case, the strong field ionization is similar to isolated atoms \cite{Zherebtsov}. Since both carbon and silicon have $p$ electrons in the outer-shell, our results show that spin-polarized electron pulses will be produced. Thus, femtosecond temporal resolution can be combined with spin-polarization, adding additional capability to  ultrafast coherent structural probes.
Second, our analysis is not restricted to noble gas atoms. Similar effects should occur in linear and ring-shaped molecules with degenerate HOMO and ground singlet state. They should yield higher degree of total integrated spin-polarization than atoms, because  spin-orbit states in such molecules have lower degeneracy  than  in atoms.  Two-fold degeneracy (e.g. Eq.\,\ref{spinpol1}) of the lowest ionic state will yield up to 100\% of total spin polarization, whereas four-fold degeneracy of lowest ionic state (e.g. Eq.\,\ref{spinpol2}) yields maximum 50 \% of total spin-polarization.
Third, relatively strong total spin polarization signal can be used to probe chiral molecules with strong fields,
extending similar capabilities of the one-photon spin polarization spectroscopy \cite{review}.

We gratefully acknowledge stimulating discussions with Prof. M.\,Ivanov and Prof. R. D\"{o}rner.
We are grateful to Prof. P. Lambropoulos for his insightful comments on spin-polarization in multiphoton regime.
This work was supported by the DFG Grant No.\ Sm 292/2-1.


\end{document}